\documentclass[9pt,twocolumn,twoside]{osajnl}

\journal{ol} 

\usepackage{color}
\definecolor{mygreen}{rgb}{0,0.5,0} 
\definecolor{myyellow}{rgb}{0.87,0.8,0.47} 
\definecolor{mymagenta}{cmyk}{0,1,0,0.12} 
\definecolor{myorange}{rgb}{.95,0.5,0} 
\definecolor{myblue}{rgb}{0,0.0,0.5} 
\newcommand{\ctext}[1]{{\color{black}#1}}
\newcommand{\gtext}[1]{{\color{black}#1}}

\setboolean{shortarticle}{true} 

\ifthenelse{\boolean{shortarticle}}{\colorlet{color2}{color2b}}{\colorlet{color2}{color2}} 

\title{\LaTeX\ template for preparing an article for submission to OSA journals \emph{Applied Optics}, \emph{Advances in Optics and Photonics}, JOSA A, JOSA B, and \emph{Optics Letters}}

\author[1,2,*]{Joanna A. Zieli\'nska}
\author[1,3]{Morgan W. Mitchell}

\affil[1]{ICFO-Institut de Ciencies Fotoniques, The Barcelona Institute of Science and Technology, 08860 Castelldefels (Barcelona), Spain}
\affil[2]{presently at: National Physical Laboratory, Teddington, Middlesex TW11 0LW, United Kingdom}
\affil[3]{ICREA-Instituci\'{o} Catalana de Recerca i Estudis Avan\c{c}ats, 08010 Barcelona, Spain}

\affil[*]{Corresponding author: joanna.zielinska@icfo.eu}

\dates{Compiled \today}

\ociscodes{(270.6570) Squeezed states; (190.4970) Parametric oscillators and amplifiers ; (190.4400) Nonlinear optics, materials}

\doi{\url{http://dx.doi.org/10.1364/ao.XX.XXXXXX}}

\title{Atom-resonant squeezed light from a tunable monolithic ppRKTP parametric amplifier
}

\begin{abstract}
We demonstrate vacuum squeezing at the $D_1$ line of atomic rubidium (\unit{795}{\nano\meter}) with a tunable, doubly-resonant, monolithic sub-threshold optical parametric oscillator in periodically-poled Rb-doped potassium titanyl phosphate. The squeezing appears to be undiminished by a strong dispersive optical nonlinearity recently observed in this material.
\end{abstract}

\setboolean{displaycopyright}{true}

\usepackage{SIunits}

\begin{document}

\maketitle
\thispagestyle{fancy}

\ifthenelse{\boolean{shortarticle}}{\ifthenelse{\boolean{singlecolumn}}{\abscontentformatted}{\abscontent}}{}

%
%
%
%
%
%



Optical squeezing, in which the noise of a property such as intensity, phase or polarization is reduced below the coherent-state level, has applications in sensing \cite{CavesPRD1981, LIGONP2011}, quantum information \cite{BraunsteinBOOK2010}, and quantum nonlocality \cite{GarciaPatronPRL2004}.  Interfacing squeezed light to resonant material systems extends these applications to atomic sensing \cite{WolfgrammPRL2010},  spectroscopy \cite{PolzikPRL1992, LuciveroPRA2016, LuciveroPRA2017} and quantum memory \cite{AppelPRL2008, HondaPRL2008}.  Squeezed light sources at the rubidium D-line wavelengths of \unit{780}{\nano\meter} and \unit{795}{\nano\meter} allow interaction with popular warm vapor, cold- and ultra-cold atomic systems. 


Optical parametric oscillators (OPOs) consisting of a second-order optical nonlinearity in a resonator, pumped below threshold by the second harmonic of the optical frequency to be squeezed, are a versatile source of squeezed light.\ctext{This method of generating squeezing naturally gives single-longitudinal-mode output, in contrast to single-pass,  waveguide and four-wave mixing schemes. Macroscopic cavity-based squeezers operate well in laboratory environments, but typically are large and require both passive vibration isolation and active cavity-length stabilization at audio frequencies. Using a resonator around the crystal also places strong demands on the crystal surface coatings, which must be low-loss for the squeezed wavelength and able to resist high intensities at the pump wavelength.  Monolithic OPOs have been developed as a response to these issues, but to date they either do not offer cavity tunability \cite{Ast13}, or tune the cavity at the cost of decreasing phase-matching \cite{Yonezawa10}. }

Monolithic OPOs \cite{Yonezawa10,Ast13}, in which a single crystal acts as both nonlinear material and optical resonator, offer important advantages in stability, size, and efficiency. More fundamentally, the absence of air-crystal interfaces in these devices reduces losses and the potential for damage by strong pump intensities, key factors in the achievable squeezing. Indeed, monolithic devices hold records for optical squeezing up to \unit{15}{dB} at wavelengths beyond \unit{1}{\micro\meter} \cite{MehmetOE2011, Ast13, VahlbruchPRL2016}.
 
Realizing high squeezing levels at atomic wavelengths remains an open challenge. The relatively short wavelengths affect the technologies used for pumping and coating, the nonlinear material itself, and phase-matching. A promising new material is periodically-poled Rb-doped potassium titanyl phosphate (ppRKTP) which is only weakly absorptive at the second harmonic of the Rb $D_1$ and $D_2$ lines, and has poling advantages relative to undoped KTP \cite{LiljestrandOE2016}. Incorporation of this material in a monolithic OPO is promising for atomic quantum optics.

Prior work has demonstrated monolithic KTP devices \cite{Yonezawa10} \ctext{that were  not fully tunable, limiting \gtext{their} applicability.} We recently reported a doubly-resonant monolithic frequency converter  in ppRKTP, with three thermal degrees of freedom enabling full tunability \cite{ZielinskaOE2017}. \gtext{To our knowledge, this is the first monolithic device to ensure cavity resonances for both pump and squeezed beams at arbitrary wavelengths, together with phase-matching.} Here we study the suitability of such devices for quantum optical applications. When used as a frequency doubler, the ppRKTP device showed an as-yet-unexplained optical nonlinearity producing strong optical bistability features \cite{Zielinskaarx2017}.  While advantageous for the tuning of the frequency doubler, the effect of this nonlinearity on quantum noise properties is unknown.  Here we demonstrate quadrature squeezing from this device, confirming its suitability for quantum optical experiments.

\begin{figure}[t]
	\centering
 \fbox{\includegraphics[width=1.05\columnwidth, trim=0.75cm 0cm 0cm 0cm]{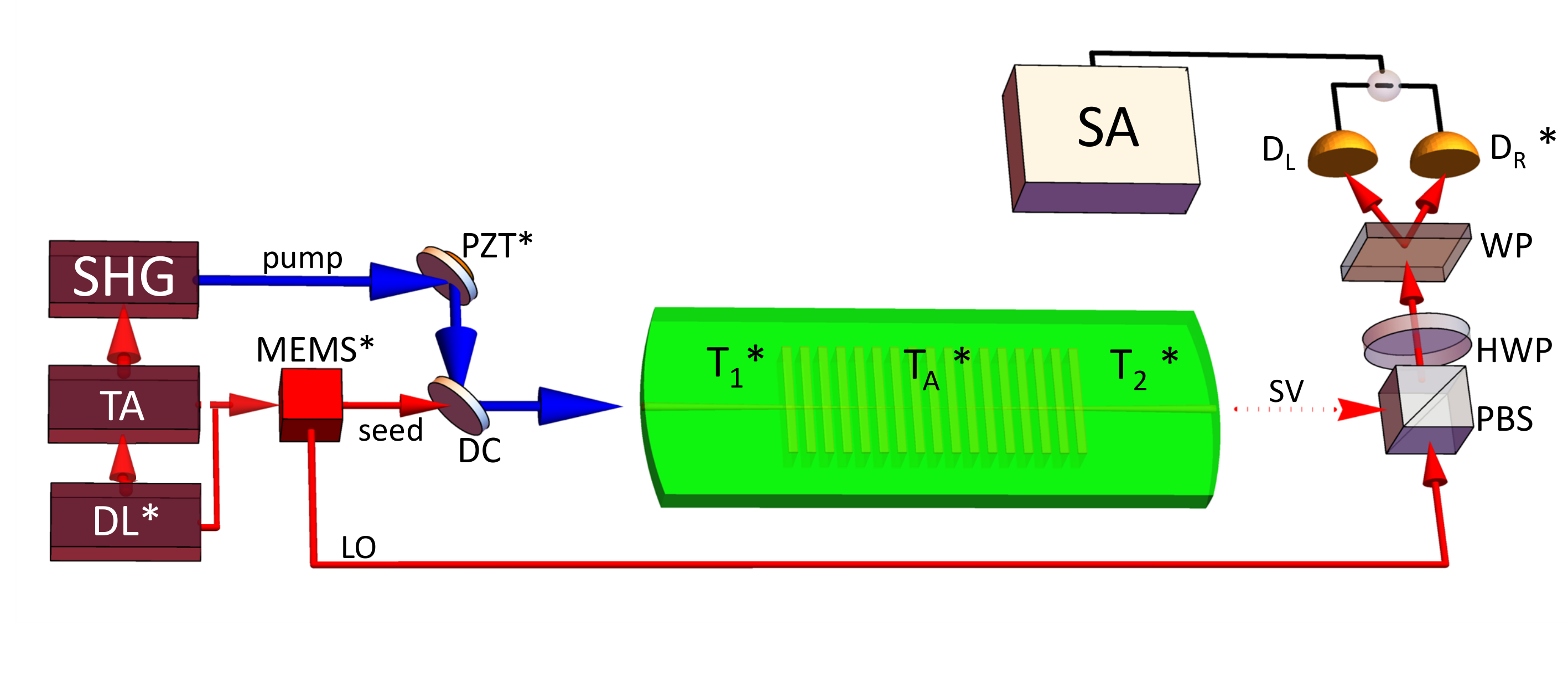}}
 \caption{Experimental setup: DL - diode laser, TA - tapered amplifier, \ctext{ SHG - second harmonic generation system,} PZT - piezoelectric actuator, MEMS - micro-electromechanical system (switching input beam between two outputs), DC - dichroic mirror, PBS - polarizing beam splitter, HWP - half wave plate, WP - Wollaston Prism, $\rm{D_R}$ and $\rm{D_L}$ - balanced detector inputs, SA - spectrum analyzer, LO - local oscillator, SV - squeezed vacuum, $T_1$ and $T_2$ - temperatures of the non-poled sides of the crystal, $T_A$ - temperature of the active section of the crystal.
 Blue and red arrows represent 397 nm  and 795 nm light respectively. Elements marked by an asterisk (*) are connected to a FPGA-based real-time control system.}
 \label{setup}
\end{figure}

The experimental setup is shown schematically in Fig.~\ref{setup}.  An external-cavity diode laser (ECDL) and tapered amplifier (TA) produce the fundamental light at {795}{\nano\meter}. Most of the TA power is frequency doubled to \unit{397}{\nano\meter} in order to generate a pump beam for the OPO, while a fraction is reserved for the local oscillator (LO) in homodyne detection and a seed beam to study the classical amplification properties of the device.  A spectrum analyzer (SA) analyzes the homodyne detection output.

Details of the doubly-resonant monolithic frequency converter and its tuning mechanisms are presented in \cite{ZielinskaOE2017}. Four degrees of freedom: the temperatures of three sections of the crystal and the pump laser frequency, are used to produce: phase matching, the fundamental and second-harmonic cavity resonance, and constructive interference between forward- and backward-emitted down-conversion light.

A microelectromechanical (MEMS) device switches fundamental power between the seed and LO beams, with only one powered at any given time. When the seed is on, the photocurrent of detector D$_R$ indicates the \unit{795}{\nano\meter} transmission of the cavity, of interest for cavity stabilization and classical gain measurements.  When the LO is on, the differential signal D$_R$-D$_L$ indicates one quadrature of the \unit{795}{\nano\meter} field emitted by the cavity, of interest for the squeezing measurements.

An FPGA-based controller sets the laser current,  three thermal degrees of freedom of the monolithic cavity, pump beam phase (via controlling the voltage fed to  a piezo-electric actuator in the pump path)  and the MEMS switch. The switching between seed and LO beams occurs at \unit{10}{\hertz}, with a duty cycle of 50$\%$. The trigger signal for the pump phase sweep is synchronized with the switching, in order to avoid injecting additional noise through the gain's phase dependence.  Due to the large resonance bandwidth of \unit{250}{\mega\hertz} and high stability of the thermally-controlled cavity, a very simple slow feedback strategy suffices to maintain the fundamental beam resonance. The FPGA notes the maximum and minimum of the seed exiting the cavity during the  $\sim 10\pi$ pump phase sweep when the LO is off and the seed is periodically amplified and deamplified due to interaction with the pump. Basing on the measurement of the maximum of seed power, the ECDL frequency is "walked" in steps of \unit{2}{\mega\hertz} every \unit{0.1}{\second}, reversing direction of the steps whenever the detected maximum seed power drops by more than 0.5 $\%$ relative to the last step. 

While only the maximum is used for ensuring the fundamental beam resonance, both maximum and minimum are necessary to calculate the gain, which is used as an indicator for the two remaining degrees of freedom of the monolithic cavity, namely pump resonance and phase between backward- and forward-interacting light. These two conditions are controlled using thermal degrees of freedom of the cavity, without losing the fundamental beam lock, by adjusting the temperatures to maximize the gain at a given pump power. \ctext{The source would remain stable for up to one hour, the factor limiting its stability being the fluctuating power of the strongly absorbed pump light.}

Finally, the homodyne detection scheme is employed using local oscillator mode-matched to the seed beam exiting the cavity with at least 98$\%$ visibility of the interference obtained on one side of the balanced detector by scanning the local oscillator phase. The gain setting of the differential output of the balanced detector corresponds to a bandwidth of \unit{45}{\mega\hertz}.


The first step in the OPO characterization was measuring the parametric gain, a phase sensitive amplification of the seed light, as a function of pump power. During this measurement the lock was continuous (seed always on), since there was no need for the local oscillator. The gain is obtained by measuring minimum power $P_{\rm min}$ (seed deamplified by pump) and maximum power $P_{\rm max}$ (seed amplified) of the seed exiting the locked cavity on the detector $\rm{D_R}$ over one period of pump phase modulation (consisting of a few minima and maxima) using the following formula \cite{PredojevicTh}.
\begin{equation}
G=\frac{1}{4}\Big(\sqrt{\frac{P_{\rm max}}{P_{\rm min}}}+1\Big)^2
\end{equation}

In order to measure maximum gain for a given pump power the central section of the crystal needs to be kept in phase matching temperature, whereas side sections $T_1$ and $T_2$ need to be adjusted in order to maximize first the blue resonance by $T_S=0.5(T_1+T_2)$ and then interference between forward and backward generated light by $T_D=T_1-T_2$. The procedure is described in detail in \cite{ZielinskaOE2017}.

\begin{figure}[htbp]
	\centering
 \fbox{\includegraphics[width=\columnwidth]{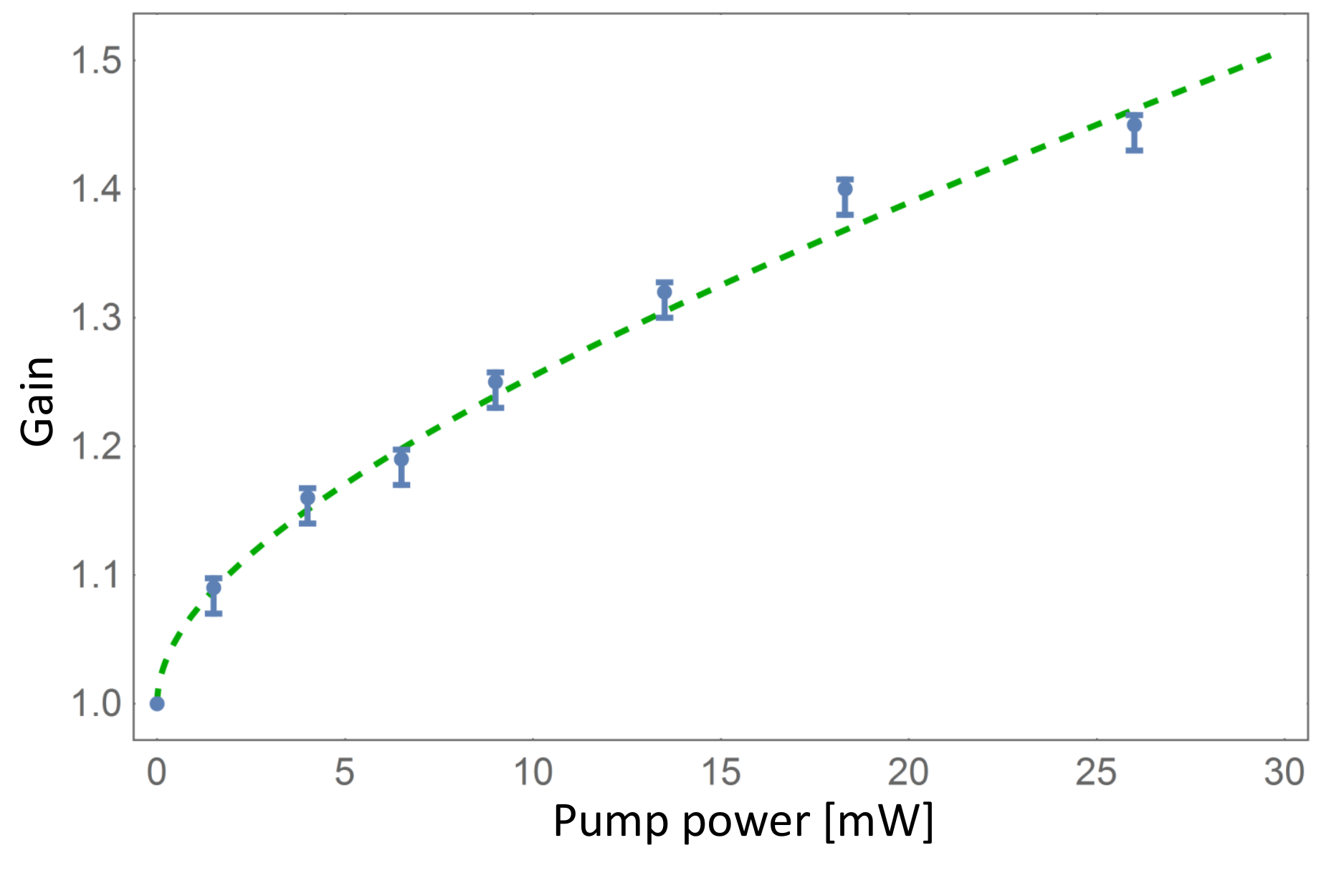}}
 \caption{Points represent gain measurement as a function of pump power, optimized with  crystal side temperatures $T_1$ and $T_2$. The green dashed line correponds to the fitted curve according to Eq. (\ref{gaineq}) with fitting parameter $P_{\rm th}=\unit{870}{\milli\watt}$. }
  \label{gain}
\end{figure}

The measurements of the optimized gain for different powers are presented in the Figure \ref{gain}. The expected relation depends on the OPO threshold pump power $P_{\rm th}$ as follows

\begin{equation}
G(P)=(1-\mu)^{-2}
\label{gaineq}
\end{equation}
where $\mu=\sqrt{{P}/{P_{\rm th}}}$ and $P$ is pump power. 

The data fit the gain vs power dependence for threshold power of 870mW (Fig. \ref{gain}). The threshold is related to the single-pass conversion efficiency $d$ by the following formula (adapted from \cite{PredojevicTh})

\begin{equation}
P_{\rm th}=\frac{T_{\rm{P}}}{1-T_{\rm{P}}}\times\frac{T^2}{4 b d}
\label{threshold}
\end{equation}
where  $T=0.14$ is red output coupler transmission, $T_{\rm{P}}=0.31$ is pump input coupler transmission and double-pass enhancement factor is defined as $b=(2-\frac{T_{\rm P}}{2})^2$. Eq. (\ref{threshold}) assumes critical coupling of the blue cavity, which means that the roundtrip loss is equal to the input coupler transmission $T_{\rm P}$, and negligible red intracavity losses. We find that single-pass efficiency yields $\unit{0.106\%}{\watt^{-1} \centi\meter^{-1}}$.


A pump-off noise measurement (meaning the difference current between $\rm{D_R}$ and $\rm{D_L}$) at the squeezing measurement conditions as a function of the local oscillator power allows us to determine the regime in which the detection system is shot noise limited, and find the electronic (independent of LO power) noise component, that will later be subtracted from the total noise in order to determine squeezing.

\begin{figure}[htbp]
	\centering
\fbox{\includegraphics[width=\columnwidth]{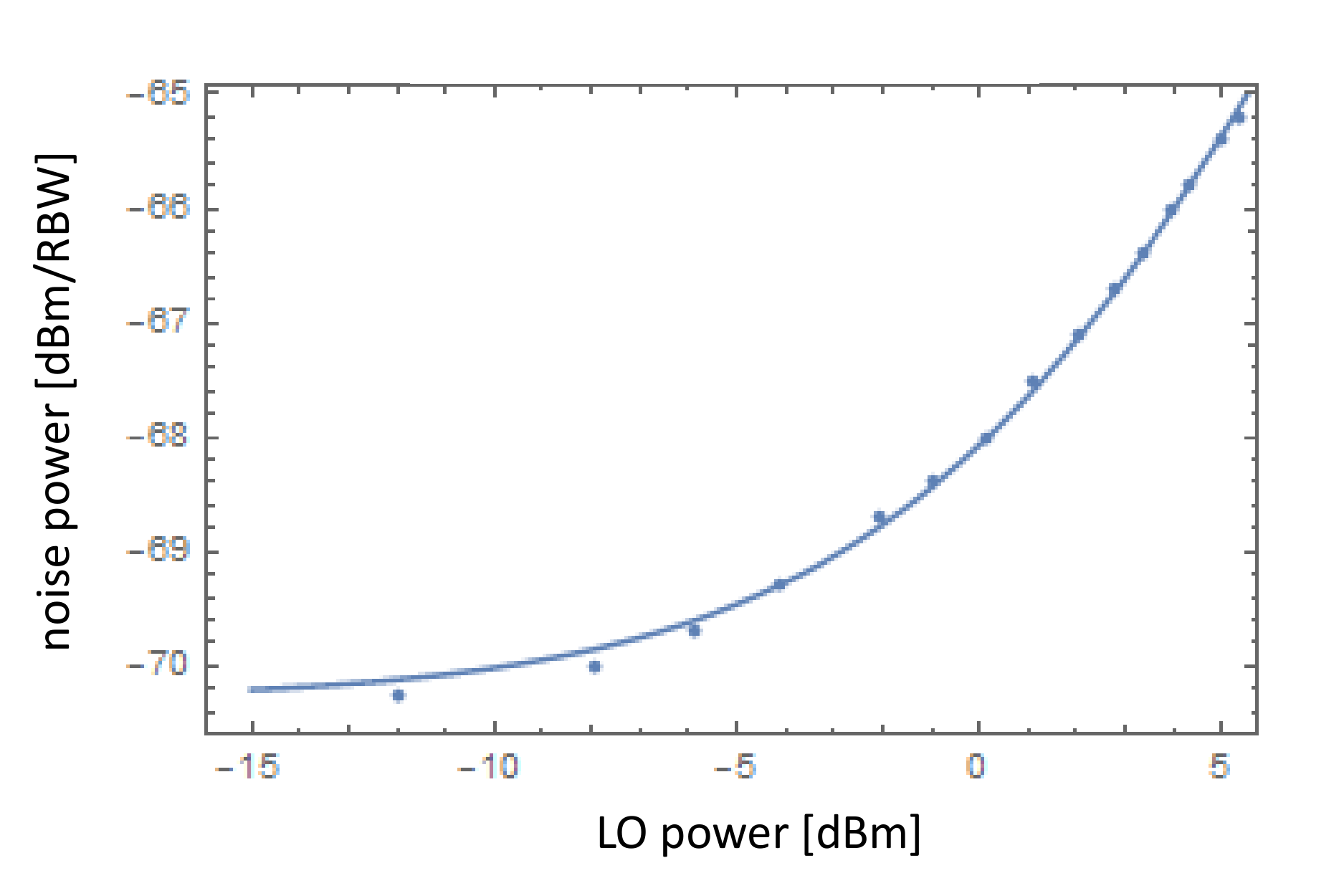}}
\caption{Points represent noise power as a function of a LO power at center frequency of 10MHz and RBW 3MHz. Solid line represents a fitted linear dependence with offset being equal to electronic noise of the system.}
\label{noise}
\end{figure}

The result of this measurement is presented in the  figure \ref{noise}. The spectrum analyzer was set to zero span mode with center frequency of 10MHz, video bandwidth (VBW) 100Hz and resolution bandwidth(RBW) 3MHz. The data are fitted with a linear function with offset that can be interpreted as electronic noise, found to be  -70.75 dBm/3MHz. From the good agreement of the linear fit with the data we deduce that our system is shot noise limited at least up to 3 mW of the LO power.
 

\begin{figure}[htbp]
	\centering
\fbox{\includegraphics[width=\columnwidth]{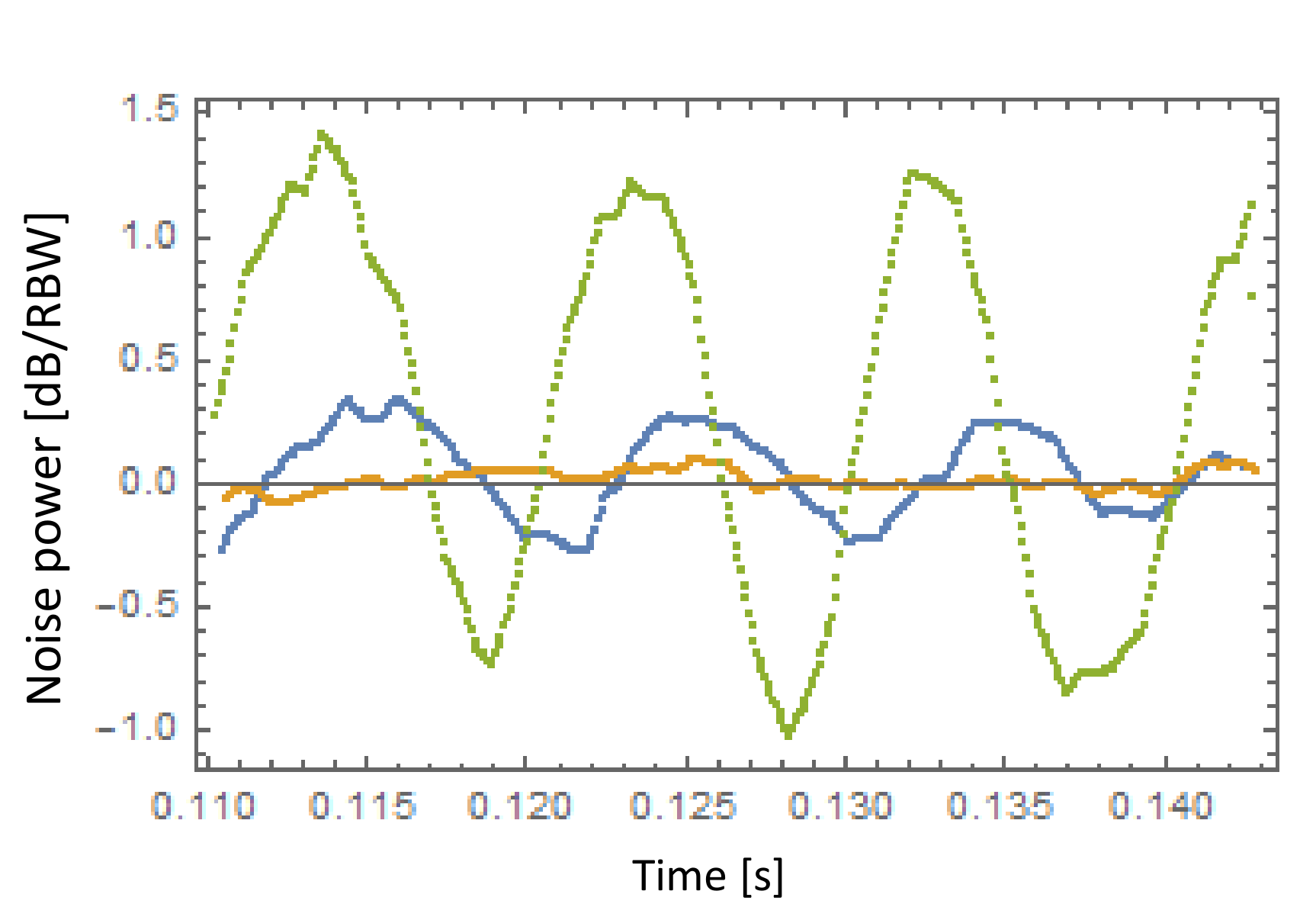}}
\caption{Green curve represents noise measured by the SA throughout the pump phase sweep with LO power of 2.5 mW and orange curve shows the same measurement with the squeezer off, which defines the zero of the vertical scale. Blue curve corresponds to the same measurement with a $50\%$ absorber inserted into the SV beam.}
\label{squeezing}
\end{figure}

Expected squeezing spectrum measured by the spectrum analyzer for $\mu<<1$ can be calculated as \cite{PredojevicTh}

\begin{equation}
S_-(\Omega)=1-\frac{4\eta\mu}{(1+\mu)^2+(\frac{\omega}{\Delta\omega})^2}
\end{equation}

where $\omega$ is the detection frequency, $\Delta \omega$  denotes the cavity bandwidth and $\eta=\eta_{\rm det}\eta_{\rm hom}^2\eta_{\rm loss}\eta_{\rm cav}$ describes the combined effect of all the losses, including cavity escape efficiency $\eta_{\rm cav}$, homodyne visibility $\eta_{\rm hom}$, detector efficiency $\eta_{\rm det}$, and other propagation losses in the squeezed beam $\eta_{\rm loss}$. 

We realistically assume $\eta=0.75$ due to 98$\%$ of homodyne visibility,
95$\%$ of propagation loss, the cavity escape efficiency of 95$\%$ and the
quantum efficiency of the detector (Thorlabs PDB450A with windows
removed) given by the manufacturer which yields approximately 90$\%$. At $\Omega \ll 1 $, we expect close to 2dB squeezing and antisqueezing for pump power corresponding to the gain value of 1.4.

We measure the squeezing and antisqueezing by modulating the phase between LO and pump beams, using the PZT (see Fig. \ref{setup}) and recording on the spectrum analyzer how the noise varies with time. 


The data obtained yield around $1$ dB of squeezing and $1.2$ dB of antisqueezing at a frequencies small compared to the cavity bandwidth (green curve in Fig. \ref{squeezing}). Subtracting the electronic noise, this result corresponds to $1.6$ dB of squeezing and $1.7$ dB of antisqueezing.

Finally, the effect of inserting a $50\%$ neutral density filter into the supposed squeezed vacuum light was recorded (see blue curve in Fig. \ref{squeezing}), in order to make sure that generated state is a squeezed vacuum state, and not amplified and deamplified light leaking through the MEMS switch into the cavity.
The insertion of the filter into the SV beam is expected to reduce the squeezing (and antisqueezing) level from 1~dB to 0.5~dB and without adding any offset in the noise vs time dependence. No significant offset noise level change is observed, and the drop in squeezing is slightly bigger than anticipated, possibly due to filter introducing small misalignment of mode-matched squeezed vacuum and LO beams.

As described in \cite{Zielinskaarx2017}, this monolithic cavity displays a strong dispersive nonlinearity, resembling a Kerr effect with a magnitude depending on the intensity history of the \unit{795}{\nano\meter} light present in the cavity, and with a time constant of approximately \unit{12}{\second}.  As this nonlinearity does not yet have a microscopic explanation, it is interesting to ask whether it is accompanied by excess noise that might impede quantum optical experiments. We find that the single-pass efficiency of $\unit{0.106\%}{\watt^{-1} \centi\meter^{-1}}$ calculated from the gain measurements is consistent with the observed squeezing. Similarly, it agrees with the conversion efficiency of the same crystal when employed as a second harmonic generation device from \unit{397}{\nano\meter} to \unit{795}{\nano\meter} (see \cite{ZielinskaOE2017}), operating as a "reversed" degenerate OPO. We conclude that, within our sensitivity of $\sim 0.1$ units of vacuum noise, there is no evidence for such excess noise in the OPO. Finally, we note that the  efficiency of the frequency converter is considerably below what is expected for an ideally-poled crystal, possibly due to poling imperfections.

We have observed 1.6 dB of squeezing from a fully-tunable doubly resonant monolithic cavity in ppRKTP. It is a key step in developing an efficient, compact, portable and vibration-insensitive source of atom resonant squeezed light for various quantum optics experiments, with the use of two crystals, one for second harmonic generation, and the second one, a squeezer, pumped by the first one (so that only one fiber input for 795 nm pump light is necessary). We show that the third order nonlinear effect previously observed in  ppRKTP \cite{Zielinskaarx2017} does not cause an increase in quantum noise, making the material suitable for squeezed light generation by spontaneous parametric downconversion.

\section*{Acknowledgements}

We thank Carlota Canalias and Andrius Zukauskas from KTH
Royal Institute of Technology in Stockholm for valuable discussions
and for fabricating the RKTP crystals.

\section*{Funding}
This work was supported by European Research Council (ERC) projects AQUMET (280169) and ERIDIAN (713682); European Union QUIC (641122); Ministerio de Economia y Competitividad (MINECO) Severo Ochoa programme (SEV-2015-0522) and projects MAQRO (Ref. FIS2015-68039-P), XPLICA  (FIS2014-62181-EXP); Ag\`{e}ncia de Gesti\'{o} d'Ajuts Universitaris i de Recerca (AGAUR) project (2014-SGR-1295); 
Fundaci\'{o} Privada CELLEX; Generalitat de Catalunya (CERCA Programme); J.Z. was supported by the FI-DGR PhD-fellowship program of the Generalitat of Catalonia.

\bibliography{squeeze}

\newpage

\newpage

\bibliographyfullrefs{squeeze}

\end{document}